\documentclass[%
 aip,
 amsmath,amssymb,
 reprint,%
]{revtex4-1}

\usepackage{graphicx}
\usepackage{dcolumn}
\usepackage{bm}

\usepackage[utf8]{inputenc}
\usepackage[T1,T2A]{fontenc}
\usepackage{mathptmx}
\usepackage{etoolbox}
\usepackage{mhchem}

\usepackage{amsmath,amsfonts,amssymb,flafter,float,graphicx,latexsym,natbib,verbatim,xfrac}
\usepackage[russian, english]{babel}
\usepackage[usenames]{color}
\usepackage[mathscr]{euscript}
\usepackage[pdftex,colorlinks=true,linkcolor=blue,urlcolor=blue,citecolor=blue]{hyperref}

\makeatletter
\def\@email#1#2{%
 \endgroup
 \patchcmd{\titleblock@produce}
  {\frontmatter@RRAPformat}
  {\frontmatter@RRAPformat{\produce@RRAP{*#1\href{mailto:#2}{#2}}}\frontmatter@RRAPformat}
  {}{}
}%
\makeatother

\DeclareUnicodeCharacter{2212}{-}

\begin{document}

\preprint{AIP/123-QED}

\title{Accuracy Bottlenecks in Impedance Spectroscopy due to Transient Effects}

\author{Victor Lopez-Richard}
\email{vlopez@df.ufscar.br}
\affiliation{Department of Physics, Federal University of São Carlos, 13565-905 São Carlos, SP, Brazil}

\author{Soumen Pradhan}
\email{soumen.pradhan@uni-wuerzburg.de}
\affiliation{Julius-Maximilians-Universität Würzburg, Physikalisches Institut and Würzburg-Dresden Cluster of Excellence ct.qmat, Lehrstuhl für Technische Physik, Am Hubland, 97074 Würzburg, Deutschland}

\author{Leonardo K. Castelano}
\affiliation{Department of Physics, Federal University of São Carlos, 13565-905 São Carlos, SP, Brazil}

\author{Rafael Schio Wengenroth Silva}
\affiliation{Department of Physics, Federal University of São Carlos, 13565-905 São Carlos, SP, Brazil}

\author{Ovidiu Lipan}
\affiliation{Department of Physics, University of Richmond, 28 Westhampton Way, Richmond, Virginia 23173, USA}

\author{Sven H\"{o}fling}
\affiliation{Julius-Maximilians-Universität Würzburg, Physikalisches Institut and Würzburg-Dresden Cluster of Excellence ct.qmat, Lehrstuhl für Technische Physik, Am Hubland, 97074 Würzburg, Deutschland}

\author{Fabian Hartmann}
\affiliation{Julius-Maximilians-Universität Würzburg, Physikalisches Institut and Würzburg-Dresden Cluster of Excellence ct.qmat, Lehrstuhl für Technische Physik, Am Hubland, 97074 Würzburg, Deutschland}

\date{\today}

\begin{abstract}
Impedance spectroscopy is vital for material characterization and assessing electrochemical device performance. It provides real-time analysis of dynamic processes such as electrode kinetics, electrons, holes or ion transport, and interfacial or defect driven phenomena. However, the technique is sensitive to experimental conditions, introducing potential variability in results. The intricate interplay of transient effects within the realm of spectral impedance analyses introduces a layer of complexity that may impede straightforward interpretations. This demands a nuanced approach for refining analytical methodologies and ensuring the fidelity of impedance characterization once the dynamic contributions of transient ingredients cannot be disentangled from the underlying steady-state characteristics. In our study, we experimentally identify that the transient effects in a memristor device are most pronounced near an optimal frequency related to intrinsic relaxation times, with these effects diminishing as the frequency varies beyond or below this range. While inherent systematic errors impose a practical limit (accuracy floor) on achievable measurement accuracy, this paper offers qualitative and quantitative insights into how specific procedures affect this limit and how to reduce it in orders of magnitude. Only by effectively addressing these errors we can push beyond this constraint.
\end{abstract}

\maketitle

\section{Introduction}

Impedance spectroscopy (IS) has emerged as an indispensable tool for material characterization and assessing the performance of electronic and electrochemical devices. In the field of material science, IS facilitates a comprehensive exploration of materials' electrical properties, offering critical insights into conductivity, dielectric response, and structural integrity.~\cite{Joshi2020} This capability is particularly crucial in the development of advanced materials for batteries,~\cite{Vadhva2021} fuel cells, solar cells,~\cite{Ebadi2019} and sensors.~\cite{PitchaiMuthusamy2024, Balasubramani2020}

When applied to electrochemical devices, IS provides real-time insights into dynamic system behavior, aiding in the optimization of performance by identifying factors such as electrode kinetics, ion transport, and interfacial processes.~\cite{Wang2021} IS plays a pivotal role by providing detailed insights into electrochemical processes at the electrode-electrolyte interface~\cite{Bredar2020} offering a comprehensive understanding of charge transfer resistance, capacitance, and diffusion processes taking place there.~\cite{Gonzales2022}

While powerful, IS is not without challenges. Sensitivity to experimental conditions introduces variability that may compromise reliability. Factors like electrode preparation, temperature fluctuations, and electrolyte nature can impact impedance spectra.~\cite{Balasubramani2020,Girija2018}
Careful interpretation is essential~\cite{Bisquert2023a,Bisquert2023b}, as misunderstanding underlying electrochemical processes or improper model selection can lead to inaccurate results.~\cite{LopezRichard2023} The choice of frequency range is critical, as overlooking relevant frequencies may result in incomplete information.

Electrochemical and electronic systems often exhibit temporal transient behaviors linked to charge transfer reactions, ion diffusion, and mass transport.~\cite{Chen2021} Understanding these time-dependent phenomena is vital for optimizing device performance~\cite{Bisquert2024}, predicting longevity, and designing efficient systems. Also, volatile memories like dynamic random-access memory involve transient charge redistribution,~\cite{Leng2022} necessitating an understanding of time-dependent processes. In turn, memristors, which are memory devices with resistance changing over time,~\cite{Paiva2022,Silva2022} exhibit dynamic behaviors as well, crucial for functions like neuromorphic abilities.~\cite{Maier2016,Kumar2020,Yang2022}

To delve deeper into the intricacies of IS for all these systems, we must critically examine the impact of inherent transients. How does impedance spectroscopy respond to and evolve amidst these transient phenomena? In what specific ways do transient interference leaves their mark on impedance spectra?

In navigating these questions, an exploration of the resilience or fragility of current measurement protocols in the face of transient effects becomes essential~\cite{Szekeres2021}. How do these protocols fare in the presence of transient influences, and can we characterize their robustness? Moreover, can we develop optimal measurement protocols ensuring more reliable and accurate impedance spectroscopy outcomes regardless of the physical origin of the transient response? Some guidelines in the literature advocate for vague recommendations, such as extending integration time or increasing the number of cycles, introducing a ``quiet time'', and disregarding the initial measured frequency due to startup transients~\cite{Lvovich2012}. However, these suggestions lack explicit clarification on the relative impact of each directive in enhancing the accuracy of the measurement results. Other guidelines propose intricate solutions involving mathematical data processing, such as differentiation~\cite{Stoynov2020}, which enhances sensitivity but demands a dense set of high-quality data, or integration~\cite{Kaisare2011}, used to apply the Kramers-Kronig relation for validating steady-state conditions. Adapted Kramers-Kronig procedures can be applied even to nonlinear systems using perturbation approaches~\cite{Hutchings1992} or by introducing phenomenological corrections based on equivalent circuit representations~\cite{Schiller2001}. Experimental compensations have also been proposed by combining IS with time-domain measurements~\cite{Popkirov1994}. In contrast, our methodological proposal eliminates the need for additional data processing.

\begin{figure}
	\includegraphics{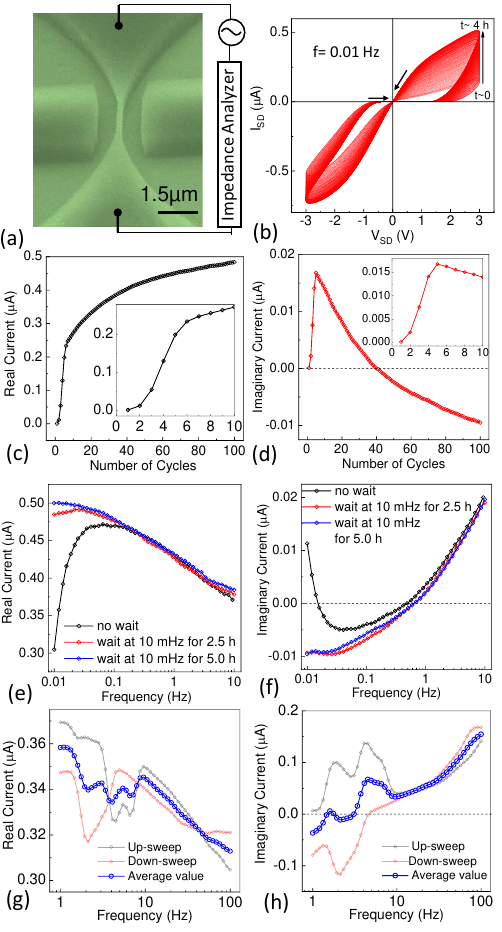}
	\caption{(a) Microscopy image of the floating gate memristive device schematically connected to the impedance spectroscopy circuitry. (b) Dynamic current-voltage response for a frequency of 0.01 Hz, and transient response of the (c) Real and (d) Imaginary parts of the current at sinusoidal voltage of amplitude 3 V and frequency 0.01 Hz. The insets depict zoomed-in images capturing the transients in current components for the first few number of cycles introduced by the instrument. (e) Real and (f) Imaginary current spectra with sinusoidal voltage of amplitude 3 V at different waiting times at starting frequency of 0.01 Hz avoiding transients induced by the instrument. The corresponding (g) Real and (h) Imaginary parts for fast integration times with a stable initial condition of 4 h wait at starting frequency for up and down-exponential sweeps and their average values.}
 \label{fig0}
\end{figure}

As we embark on this exploration, the ultimate question emerges: Can we contribute more precise guidelines than those currently available? Through a meticulous examination of these questions, we aim to advance not only our understanding of impedance spectroscopy but also our ability to harness its full potential by refining measurement protocols in the presence of transient complexities.

\section{Results and Discussion}

The experimental findings presented in Figure~\ref{fig0} offer an illustrative example to underscore and warn against the impact of transient dynamics on spectroscopic accuracy. The figure showcases the impedance analysis of a floating gate based memristive device (described in Ref.~\citenum{Miller2021}), as depicted in panel (a) by its microscopy image schematically connected to the impedance spectroscopy circuitry. At first, the current response was modulated into a stabilizing memristive loop under a periodic sinusoidal voltage drive of amplitude 3 V and frequency 0.01 Hz as shown in panel (b). It has been observed that as the number of voltage cycles increases, there is a gradual and slow saturation of the current amplitude. This observation is consistent with expectations across various conductive devices and processes~\cite{Silva2022}, including Faradaic electrochemical reactions at electrodes~\cite{Kaisare2011}, carrier thermal ionization from defects in semiconductor oxides~\cite{Paiva2022}, floating gate charging in quantum dot transistors~\cite{Gopfert2010}, and the impact of moisture on anionic–electronic resistive switches~\cite{Messerschmitt2015}. Therefore, it is highly desirable to investigate intrinsic carrier dynamics through IS analysis. The detailed experimental procedure of IS measurements are discussed in the Experimental Method section. Here, the real and imaginary components of the current response are observed in sequential measurements at a fixed voltage of frequency $f=0.01$ Hz and amplitude 3 V, as shown in panels (c) and (d), respectively. In the first few cycles, a drastic change is observed in both the real and imaginary parts as can be seen in the insets. These transients in the measurements arise from the interference of the instrument low pass filter. Within this frequency range, there is high chance of mixing noise signals close to the actual signal frequency. Currently, the sole approach to mitigate the inclusion of noise signals is by enhancing the roll-off steepness of the low-pass filter. It is evident that a higher filter order leads to a steeper roll-off and the drawback of using higher filter order is that it increases the settling time to reach the final value~\cite{SR,Zurich}. In our experiment, we have used a maximum possible filter order of 4 with roll-off of 24 dB/oct. Therefore, it is expected that the Lock-In amplifier will require the initial few voltage cycles to attain a stabilized response. Yet, instead of stabilization, the variation in both the real and imaginary current shows a gradual saturation after the 5th or 6th cycle, pointing now to the retarded dynamics of charge carriers within the device itself. The observation of transients in the device response can be ascribed to the presence of long-term relaxation processes~\cite{Paiva2022,Miller2021}. This same effect contributes to discrepancies in the spectral analysis of the corresponding real and imaginary components of the current, illustrated in panels (e) and (f). Note the gradual convergence towards a stable spectrum as the waiting time increases (red and blue symbols). In this case, insufficient waiting times could lead to misleading interpretations, particularly at lower frequencies. This may give the appearance of a false capacitive (or negative inductive) interference, which later shifts into an apparent inductive response at intermediate values before reverting back to an apparent capacitive behavior. This pattern is evident in the nonmonotonic black symbols shown in Figures~\ref{fig0} (e) and (f). A sharper contrast in spectral results can be achieved by using a fast integration time, starting from a stabilized response at either low or high frequency for upward or downward frequency sweeps, respectively. This is illustrated in Figures~\ref{fig0} (g) and (h). It appears that the sweep direction causes opposite deviations from the expected monotonic trends in panels (e) and (f). Additionally, the deviations seem more complex for the upward sweep. These phenomena are widespread and will be systematically discussed in the following.


When the electrical output changes over time and the response is a function of both the input and the current state of the system at any given moment, the system is called dynamic. In the case of conductance fluctuations, one may assume that this results in a modulation around certain equilibrium value, $G_0$, by changes of the number of available carriers, $\delta n$, no matter their nature, as
\begin{equation}
    G\left( t \right)=G_0+ \gamma \delta n\left( t \right),
    \label{Drude}
\end{equation}
where $\gamma=\frac{e \mu}{L^2}$, $e$ is the elementary charge, $\mu$ the mobility, and $L$ the distance between contacts. Additionally, the response that retains information of previous system states, within the relaxation time, $\tau$, approximation, can be expressed as the convolution
\begin{equation}
    \delta n\left( t \right)=\int_0^t g(t-t')e^{-\frac{t'}{\tau}} dt' +\delta n_0 e^{-\frac{t}{\tau}},
    \label{conv}
\end{equation}
under a generation (or trapping) function $g(t)$ and initial condition $\delta n\left( 0 \right)=\delta n_0$. In the presence of a periodic input with frequency $\omega=2 \pi f$, this leads to a solution that can be described, in general terms, as
\begin{equation}
    \delta n(t)\equiv\delta n(\omega,t)=\left[ \delta n_0 - F\left( \omega,0\right)\right] \exp\left({-\frac{t}{\tau}}\right)+F\left( \omega,t\right),
    \label{dn1}
\end{equation}
In the limit of large enough times, the transient element, proportional to $\exp(-t/\tau)$, decays exponentially over time and the expression in Eq.~\ref{dn1} converges to the periodic term, $\delta n_{\infty}=F\left( \omega,t\right)=F\left( \omega,t+\frac{2 \pi}{\omega}\right)$.
Analytical expressions for this function can be derived in a concise form, not detailed here, by expanding $g(t)$ in powers of the applied voltage, $V(t)$, according to the pulse shape.~\cite{Silva2022,Paiva2022,LopezRichard2022} The concepts of admittance or impedance and their spectral response are meaningful in the context of this stable periodic electric response, often associated with sinusoidal waveforms, where the current can be described by a periodic function as, $I_{\infty}=\left( G_0+\gamma \delta n_{\infty} \right) V(t)$, that can be decomposed in a multimode expression as~\cite{LopezRichard2023}
\begin{equation}
   I_{\infty}=V_0 \sum_n \left(G_\infty^{(n)}\cos n\omega t -B_\infty^{(n)} \sin n\omega t  \right),
\end{equation}
given the dynamic nature of the response embedded in Eq.~\ref{conv}. This definition allows not being constrained by the restriction of linearity of the response~\cite{Szekeres2021} that is clearly, according to Ref.~\citenum{LopezRichard2023}, not compulsory for impedance spectral analysis. The Fourier coefficients $G_\infty^{(n)}$ and $B_\infty^{(n)}$ can be recognized as the conductance and susceptance per mode, $n=0,1,2...$~\cite{LopezRichard2023}. However, the condition of a perfectly stable periodic response cannot be guaranteed absolutely in real-world electrical systems. 

Spectral analysis using Fourier decomposition of the response, over $m$-periods, defined as
\begin{equation}
   G^{(n)}=\frac{\omega}{m \pi}\int_0^{\frac{2\pi m}{\omega}} \frac{I(t)}{V_0} \cos( n\omega t) dt=G_\infty^{(n)}+\Delta G^{(n)},
   \label{int1}
\end{equation}
(for $G^{(0)}$ the expression must be multiplied by $1/2$) and
\begin{equation}
   B^{(n)}=-\frac{\omega}{m \pi}\int_0^{\frac{2\pi m}{\omega}} \frac{I(t)}{V_0} \sin( n\omega t) dt=B_\infty^{(n)}+\Delta B^{(n)},
    \label{int2}
\end{equation}
will unavoidably reveal divergences, labeled as $\Delta G^{(n)}$ and $\Delta B^{(n)}$, from the stable values, $G_\infty^{(n)}$ and $B_\infty^{(n)}$, and these divergences are dependent on the experimental procedure itself. They will determine the inexorable accuracy limit of any given impedance spectroscopic analysis, labeled in what follows as accuracy floor. It must be lowered in order to improve the reliability of a given IS analysis.
\begin{figure}
	\includegraphics{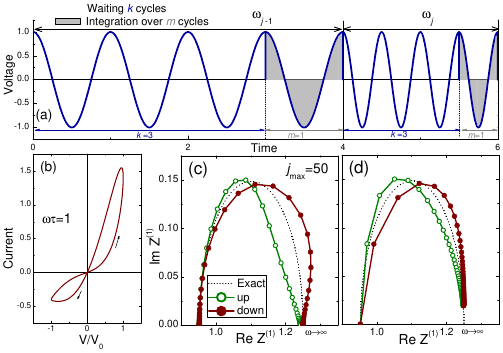}
	\caption{ (a) Depiction of two voltage segments over time, corresponding to consecutive frequencies employed for excitation, incorporating $k$ waiting periods per frequency value. The same function serves as reference for signal integration, spanning $m$ cycles (shaded region). This picture, in particular, corresponds to $k=3$ and $m=1$. (b) Current-voltage characteristic of an asymmetric memristive response for $\omega=1/\tau$. (c) The corresponding Nyquist map for the first mode: the exact, stabilized, spectrum is represented with dotted line while up and down exponential frequency sweeps are plotted with open and solid symbols, respectively. (d) The Nyquist map for linear frequency sweeps using the same patterns of panel (c). In panels (c) and (d) $j_{max}$ represents the number of sampling points in the frequency spectrum.}
 \label{fig1}
\end{figure}
Let us assume a sequence of a frequency sweep, $\omega_j$, with $j=0,1,2..., j_{max}$, for now arbitrary, that starts from an already stable condition at $\omega_0$, as suggested in methodological texts \cite{Lvovich2012} by discarding the first cycles. The literature also often suggests introducing a delay of one or two cycles per frequency before each impedance measurement to mitigate errors arising from potential transient responses during frequency transitions. However, quantitative support for the efficacy of this recommendation is notably absent~\cite{Wang2021}. Therefore, to qualify and quantify these recipes, we will simulate a sweeping voltage pulse, serving both as excitation and reference for signal integration, as illustrated in Figure~\ref{fig1} (a), consisting in $k=0$, 1,... waiting cycles, and $m=1$, 2 ... integration cycles for each frequency value, $\omega_j$. Note that we have ensured continuity in both the voltage drive and its derivative throughout the entire dynamic process. For the purpose of generating a continuous pulse in time, $t \in [0, \infty)$, with fixed frequencies $\omega_j$ across $(k+m)$ cycles in Figure~\ref{fig1} (a), the voltage should be $V(t)=V_0\cos(\omega_j t+\phi_j)$, with a discontinuous phase $\phi_j=-2\pi (k+m)\omega_j s_j$ and $s_j=\sum_{l=0}^{j-1}\omega_l^{-1}$, where $\omega_{-1}$ is considered as $\infty$ to appropriately define the initial cycle. Once the continuity conditions are set, an alternative representation for this function emerges, segmented according to each frequency value across $(k+m)$ cycles, expressed as $V(t) = V_0 \cos(\omega_j t)$. Here, time is defined exclusively within each interval $t \in [0, 2\pi (k+m)/\omega_{j}]$.

By quantifying the oscillation dependence, we develop a predictive tool that guides researchers in choosing the optimal oscillation number for maximizing experiment success and validating the proposed recommendation. We can show how the divergences depend on the number of full oscillations, $k$, discarded before the integration over $m$-cycles of Eqs.~\ref{int1} and~\ref{int2} is taken and can be analytically obtained, once the solution of Eq.~\ref{dn1} is known, as
\begin{equation}
   \Delta G^{(n)}(\omega_j)=L\left(\omega_{j}, \omega_{j-1}\right)\exp\left(-\frac{2 \pi k}{\omega_j \tau}   \right)C^{(n,m)}(\omega_j),
   \label{DGn}
\end{equation}
and 
\begin{equation}
   \Delta B^{(n)}(\omega_j)=-L\left(\omega_{j}, \omega_{j-1}\right)\exp\left(-\frac{2 \pi k}{\omega_j \tau}   \right)S^{(n,m)}(\omega_j),
   \label{DBn}
\end{equation}
where 
\begin{eqnarray}
   C^{(n,m)}(\omega)&=&\frac{\omega}{m \pi}\int_0^\frac{2 \pi m}{\omega} e^{-t/\tau} \cos(\omega t) \cos(n \omega t) dt  \\
   &=&\frac{\left( 1-e^{-\frac{2 \pi m}{\omega \tau}} \right)}{m \pi}\frac{\omega\tau \left[ (\omega \tau)^2 (n^2+1) +1 \right]}{1+(\omega \tau)^4 (n^2-1)^2+ 2(\omega \tau)^2(n^2+1) }, \nonumber
\end{eqnarray}
and
\begin{eqnarray}
   S^{(n,m)}(\omega)&=&\frac{\omega}{m \pi}\int_0^\frac{2 \pi m}{\omega} e^{-t/\tau} \cos(\omega t) \sin(n \omega t) dt  \\
   &=&\frac{\left( 1-e^{-\frac{2 \pi m}{\omega \tau}} \right)}{m \pi}\frac{ (\omega \tau)^2 n \left[ (\omega \tau)^2 (n^2-1) +1 \right]}{1+(\omega \tau)^4 (n^2-1)^2+ 2(\omega \tau)^2(n^2+1) }. \nonumber
\end{eqnarray}
While the memory of the sequential frequency sweep through preceding states is encapsulated within the expression 
\begin{equation}
    L\left(\omega_{j}, \omega_{j-1}\right)=\gamma\left[ \delta n \left( \omega_{j-1}, \frac{2 \pi (k+m)}{\omega_{j-1}} \right) - F\left( \omega_{j},0 \right)\right].
    \label{memory}
\end{equation}
Starting from a stable condition, under input of voltage frequency $\omega_{0}$, when
$\delta n \left( \omega_{0}, \frac{2 \pi (k+m)}{\omega_{0}} \right)=F(\omega_0,0)$, this effect is cumulative, although not necessarily monotonic as proven below. Finally, the values of $\delta n \left( \omega_{j}, \frac{2 \pi (k+m)}{\omega_{j}} \right)$ can be obtained in a recursive analytical form, independently on the way the frequency values are sampled, as
\begin{eqnarray}
    \delta n \left( \omega_{j}, t \right)&=&\left[ \delta n \left( \omega_{j-1}, \frac{2 \pi (k+m)}{\omega_{j-1}} \right) - F\left( \omega_j,0\right)\right] e^{-\frac{t}{\tau}} \nonumber \\
    &+& F\left( \omega_j,t\right).
    \label{dn2}
\end{eqnarray}
In exploring the dilemma of impedance analyses under non-stabilized electric responses, we will consider the case of a memristive response, reported in Ref.~\citenum{LopezRichard2022} which is based on the Drude-like conductance model described in Eq.~\ref{Drude}. Its current-voltage characteristic has been displayed in Figure~\ref{fig1} (b), as an illustrative example. While the outcomes derived from Eqs.~\ref{DGn}-\ref{dn2} can be provided for any mode order, our analysis will center specifically on the fundamental mode, $n=1$, a focal point pursued in most spectroscopic studies.

The Nyquist maps illustrating fastest sweep with zero waiting cycle and a single integration period, corresponding to $k=0$ and $m=1$ respectively, for the mode $n=1$ are depicted in Figures~\ref{fig1} (c) and (d) and have been calculated by defining the impedance as \cite{LopezRichard2023} 
\begin{equation}
Z^{(1)}=\frac{ G^{(1)}-i B^{(1)}  }{ [G^{(1)}]^2+[B^{(1)}]^2 }.
\end{equation}
In Figure~\ref{fig1} (c), the frequency has been varied within a range $[\omega_0,\omega_{j_{max}}]$, sampling $j_{max}$ values, using an exponential sweeping, $\omega_j=\omega_0 10^{\Delta j}$, with $\Delta\equiv \frac{1}{j_{max}}\log_{10} R$, and $R\equiv \frac{\omega_{j_{max}}}{\omega_0}$. In turn, the panel (d) corresponds to a linear frequency sampling with $\omega_j=\omega_0 [1+\frac{j}{j_{max}} (R-1)]$. The up and down frequency sweeps correspond to $R>1$ and $R<1$, respectively. Note, in both cases, that there will be an excess or deficiency in the absolute value of the impedance with respect to the stable case, that has been represented through the dotted circle. The larger discrepancies occur mostly at intermediary frequency values, even if starting from a perfectly stabilized condition.
\begin{figure}
\includegraphics{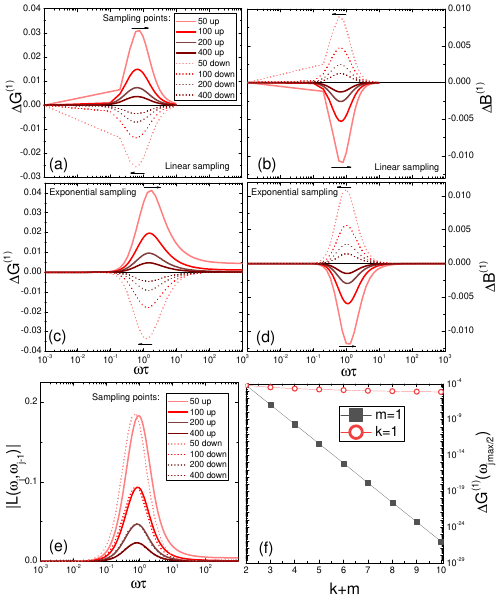}
	\caption{ Divergence as a function of driving frequency for various numbers of sampling values: (a) first mode conductance and (b) susceptance with linear sampling; (c) first mode conductance and (d) susceptance with exponential sampling. (e) Memory function, $L(\omega_j,\omega_{j-1})$, as a function of the driving frequency for up and down-sweeps and various numbers of sampling values. Evolution of the conductance divergence for $\omega_{j_{max}/2}$ as a function of either the number of integration cycles $m$ with $k=1$ or the number of waiting cycles, $k$ with $m=1$.}
 \label{fig2}
\end{figure}

This effect can be better quantified by assessing the dependence of both $\Delta G^{(1)}$, and, $\Delta B^{(1)}$ on frequency, as displayed in Figures~\ref{fig2} (a,b) and ~\ref{fig2} (c,d) for linear and exponential sampling, respectively. Note that the extent of divergence is intricately tied to the number of sampling points, $j_{max}$. Elevating the value of $j_{max}$ corresponds to a decrease in divergence. However, it is essential to recognize that this effect is not solely governed by procedural parameters. The maximum divergence consistently occurs in proximity to the resonant condition, specifically when $\omega\sim 1/\tau$. This connection links the effect to an intrinsic parameter, precisely at the condition when interesting physics unfolds. In the case of memory devices or electrochemical memristive responses, this corresponds to the condition of maximal loop areas~\cite{Silva2022}. Note also in Figures~\ref{fig2} (a-d) that the divergences for up and down-sweeps are systematically opposite in sign and almost symmetric, with the up-sweep value slightly surpassing that of the down-sweep. Thus, in principle, they can be approximately compensated by averaging the admittance values of both paths as illustrated in Figures~\ref{fig0} (g) and (h).

The relation of $\Delta G^{(1)}$ and $\Delta B^{(1)}$ and the number of sampling points, $j_{max}$, is exclusively determined by the memory function $L\left(\omega_{j}, \omega_{j-1}\right)$ in Eq.~\ref{memory}, which has been plotted in Figure~\ref{fig2} (e). It reaches its maximum at $\omega\sim 1/\tau$ and characterizes how narrowing frequency steps effectively enhances the proximity of oscillatory states, consequently mitigating the interference caused by transients. Increasing the number of integration periods, $m$, also reduces the divergence values, as illustrated in Figure~\ref{fig2} (f).
This behavior can be approximated as an inverse relation with $m$ since $C^{(1,m)}$ and $S^{(1,m)}\propto \left( 1-e^{-\frac{2 \pi m}{\omega \tau}} \right)/m$ that scales as $\sim 1/m$ in Eqs.~\ref{DGn} and~\ref{DBn} for intermediary frequencies.
This dependency contrasts with the more pronounced impact of increasing the number of waiting cycles, $k$, also depicted in Figure~\ref{fig2} (f). In that case, the exponential reduction is governed by the factor, $\exp\left(-\frac{2 \pi k}{\omega_j \tau}   \right)$, in Eqs.~\ref{DGn} and~\ref{DBn}. 
It is crucial to note that despite this uneven influence, both $k$ and $m$ carry equal weight in determining the measurement time given by
\begin{equation}
T_m=2 \pi (k+m)\sum_{j=0}^{j_{max}}\frac{1}{\omega_j}.
\end{equation}
\begin{figure}
	\includegraphics{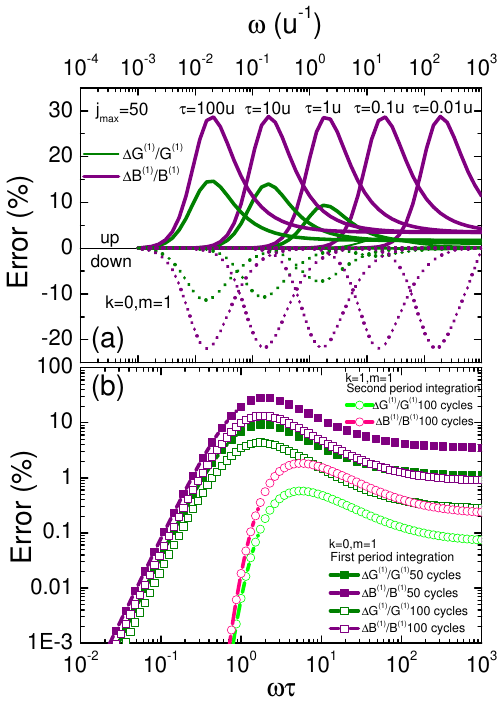}
	\caption{ (a) Calculated systematic error for the conductance and susceptance of the first mode for $k=0$ and $m=1$ as a function of the driving frequency for various values of the relaxation time, independent on the time units, $u$, including up and down exponential frequency sweeps. (b) Contrasting systematic errors for the conductance and susceptance of the first mode as function of frequency using different sampling procedures: the first period integration curves correspond to $k=0$ and $m=1$, while the second period integration curves correspond to $k=1$ and $m=1$.}
 \label{fig5}
\end{figure}
As stated previously, the systematic errors inherent in the measurement methodology are affected by intrinsic parameters as well. This has been quantified by their relative values, $\Delta G^{(1)}/G^{(1)}$ and $\Delta B^{(1)}/B^{(1)}$, in Figure~\ref{fig5} (a). In this case, it is apparent that, beside the maximum shifting, the accuracy floor for the susceptance is invariable to changes of the system relaxation time, yet the one for the conductance shrinks as the relaxation time decreases. As pointed previously, the accuracy floor is larger for up-sweeps than for the down-sweeps.

It is also important to assess the comparative effects of discarding oscillations prior to initiating the Fourier analysis integration procedure versus increasing the number of sampling points. Previously we have considered the case of zero discarded cycles, that corresponds to $k=0$ in Eqs.~\ref{DGn}-\ref{dn2}. Increasing $k$ allows for reducing the accuracy floor as represented in Figure~\ref{fig5} (b) where the case for $k=1$ is compared with the relative effect of increasing the number of sampling points. The case corresponding to $k=0$ and $m=1$ has been referred in panel  ~\ref{fig5} (b) as the first period integration while the second period integration curves correspond to $k=1$ and $m=1$. Note that dropping a single cycle per frequency doubles the measurement time in case of $m=1$ but significantly reduces, in at least one order, the accuracy floor, turning this procedure more efficient. 

\begin{figure}
\includegraphics{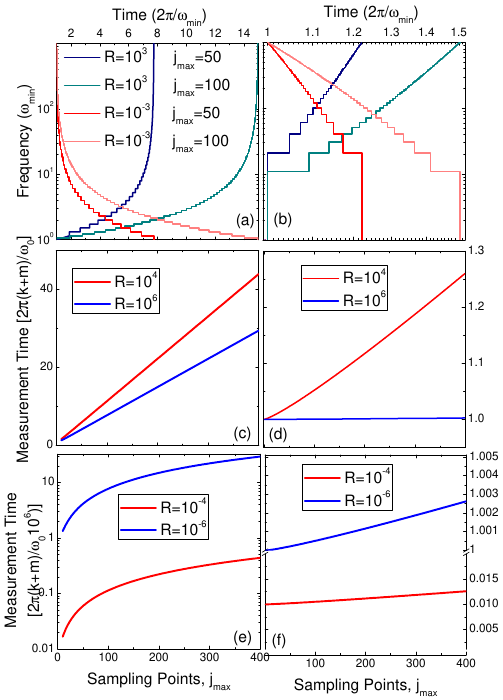}
	\caption{Frequency paths for exponential (a) and linear sweepings (b) and various sets of parameters as a function of time for a fixed minimum frequency value $\omega_{min}$. The total measurement time as a function of the number of sampling points, $j_{max}$, is represented for up-sweeps under exponential (c) and linear (d) sampling and for down-sweeps for exponential (e) and linear sampling (f), respectively.}
 \label{fig4}
\end{figure}

The choice between linear and exponential frequency sweeps depends on the specific characteristics of the system under investigation and the goals of the impedance spectroscopy experiment. The linear frequency sweeps in Figure~\ref{fig2} (a,b) yield smaller divergences than the exponential sweeps in Figure~\ref{fig2} (c,d), thus the pros and cons of each procedure must be assessed and weighted. The frequency sampling of potential experimental paths has been represented as a function of time in Figures~\ref{fig4} (a) and (b) for exponential and linear sweeps, respectively. 
Exponential sweeps, in panel (a), provide higher resolution at lower frequencies, allowing for better characterization of processes occurring at longer timescales, and offer increased sensitivity to changes in certain frequency regions as already shown in Figure~\ref{fig2}. Linear sweeps, in turn, as represented in Figure~\ref{fig4} (b) may result in uneven resolution in specific regions, with limited sensitivity. 
An exponential frequency sweep achieves better resolution at lower frequencies because it allocates more sampling points to this region, allowing for more detailed data collection where the signal changes more gradually. Conversely, a linear frequency sweep provides better resolution at higher frequencies by distributing sampling points more evenly across the frequency range, which is advantageous for capturing rapid changes in the signal at these frequencies. Thus, the choice of sweep type allows for optimized resolution tailored to the frequency range of interest.

Figures \ref{fig4} (a) and (b) additionally highlight the asymmetries observed in both down and up-sweeps when maintaining a constant value of $|R|$. Specifically, there is a discrepancy in the time spent at each frequency value, with a reduction during up-sweeps and an increase during down-sweeps. This results in a frequency shift of the maximum of the function $L(\omega_j,\omega_{j-1})$ according to the sampling direction, displayed in Figure~\ref{fig2} (e). In turn, this asymmetry manifests in Figures~\ref{fig2} (a)-(d), where a smaller divergence is evident for the latter case. Please note that the divergence value is influenced by the weighting function, $\exp(-\frac{2\pi k}{\omega_j \tau})$, which introduces uneven contributions depending on the region where the function $L(\omega_j,\omega_{j-1})$ peaks.

The total measurement time depends on sampling parameters. Although increasing the number of sampling points is clearly a recipe for reducing all the divergences, one of its primary drawbacks is the increase in measurement times. This can be a critical factor in time-sensitive applications or when efficiency is a priority. Extended measurement times may also introduce practical challenges related to resource utilization and experiment planning. This fact has been characterized in Figures~\ref{fig4} (c) and (d) for up-sweeps under exponential and linear samplings, respectively, and for the corresponding down-sweeps, represented in Figures~\ref{fig4} (e) and (f).

Note that the total time increase is almost linear with $j_{max}$ in all cases. We have excluded from this count the settling time needed to guarantee convergence at the initial frequency, $\omega_0$, that should be larger than $\tau$. To avoid working blindly, it is imperative to establish the value of $\tau$ beforehand. This can be achieved either by employing transient measurements, as proposed in Ref.~\citenum{LopezRichard2023}, or by verifying convergence, as depicted in Figures~\ref{fig0} (c) and (d). For the exponential sampling,  
\begin{eqnarray}
T_m &=& \frac{2 \pi (k+m)}{\omega_0}
\frac{\left[ 1-10^{-W \left( \frac{j_{max}+1}{j_{max}}\right)} \right]}{\left[ 1-10^{-\left( \frac{W}{j_{max}}\right)} \right]  } \nonumber \\
&\simeq & \frac{2 \pi (k+m)}{\omega_0} \left[\frac{1-10^{-W}}{W \ln 10}j_{max}+C\right].
\end{eqnarray}
Note that the last approximation is valid for large enough $j_{max}$, where $W=\log_{10} R$ and $C=(1-10^{-W})/2$, showing a linear behavior. For the linear sampling, the dependence is not that trivial since
\begin{equation}
T_m=\frac{2 \pi (k+m)}{\omega_0}j_{max} \sum_{j=0}^{j_{max}} \frac{1}{j_{max} +(R-1)j},
\end{equation}
that cannot be analytically reduced. In Figures~\ref{fig4}~(c)-(f), it is noteworthy that, under identical parameter settings, linear frequency sweeps consistently yield the fastest measurement procedure.

\section{Conclusions}

In summary, a judicious application of IS, coupled with careful experimental design and data analysis, remains indispensable for gaining valuable insights into electrochemical systems and conductive materials. Here, we proved and quantified the existence of unavoidable accuracy floors due to inherent non-equilibrium dynamics that cannot be overcome without addressing and mitigating the underlying systematic issues. We have unambiguously proved that this can be done by one or a combination of the following procedural ways:
 \begin{itemize}
 \item Increasing sampling points, while lowering the accuracy floor, is less efficient due to its proportional increase in measurement time.
 \item Opt for down frequency sweeps instead of up-sweeps. 
  \item Averaging the admittance values for the up and down frequency sweeps, doubling the measurement time, but almost canceling out opposite divergences for all modes.
  \item Enhancing the number of waiting cycles, $k$, per sampled frequency proves more effective in reducing the accuracy floor compared to increasing the number of integration periods, $m$, while maintaining equivalent weights in the measurement time, proportional to $(k+m)$. Nevertheless, increasing $m$ may be pertinent in mitigating spurious contributions arising from random fluctuations and noise that tend to be averaged down this way.
  \item At low frequencies, it is essential to account for adequate waiting times for both the instrument and the device to achieve stabilized values. Thus,  acquiring in advance insights into relaxation time scales by using alternative transient measurements and into the instrument time constants provides straightforward indications of the required settling periods.
  \item The relaxation time of the measured device must be assessed in advance and factored adequately into the data acquisition process, as pointed above.
\end{itemize}
Addressing and quantifying the systematic errors introduced by transient phenomena will certainly contribute to an improved experience when using IS tools, facilitating more accurate and reliable measurements. Our method's principles, rooted in the presence of relaxation time toward equilibrium, can be extended to any transport response with relaxation or reactive components, and while derived for drift transport, they can also be applied to displacement components, yielding similar—though not identical—results. We believe that these suggested enhancements will benefit both novice and experienced users, fostering a more intuitive, efficient planing and operation of IS protocols.

\section{Methods}

The dynamic current-voltage measurements were performed on the device with floating gate condition in memristor configuration~\cite{Miller2021} using sinusoidal voltage of frequency 0.01 Hz up to $\sim$4 h. The impedance spectroscopy measurements were carried out using a Lock-In amplifier (EG\&G Instruments, Model: 7265) in current signal mode. At first, the `Auto Phase' operation was executed with a standard signal to nullify the inductive or capacitive effect of connected wires or any other cross-talk in the circuit. We have performed this with DC coupling using a function generator and the connection wires, without connecting the device. Then, the real and imaginary parts of the current were measured on the actual device with floating gate configuration using a sinusoidal signal from the function generator and utilizing a same signal as reference for the Lock-In. To measure the transient response at 0.01 Hz, we configured the Lock-In amplifier with a time constant (TC) of 100 s and a roll-off of 24 dB/oct. Both the real and imaginary currents were recorded during each voltage cycle. Subsequently, the real and imaginary current spectra were acquired over the frequency range of 0.01 - 10 Hz, maintaining the same TC and roll-off settings. The data acquisition time was set to 1200 s ($12 \times$ TC) to minimize transient interference from the Lock-In amplifier. This measurement was repeated after waiting periods of 2.5 h and 5 h at starting frequency. Similarly, a fast integration of real and imaginary current spectra were recorded in the range of 1 - 100 Hz for both the up and down frequency sweeps after a stable initial condition, attained after 4 h wait at starting frequency. Here, fast integration time of 1 s was utilized at each point in the frequency spectrum with a Lock-In amplifier TC of 1 s, which is 1 cycle time of the lowest frequency, 1 Hz.

\section{Acknowledgments}

This study was financed in part by the Conselho Nacional de Desenvolvimento Científico e Tecnológico - Brazil (CNPq) Proc. 311536/2022-0, FAPESP Proc. 2023/05436-3, and the Fulbright Program of the United States Department of State’s Bureau of Educational and Cultural Affairs. The authors extend their sincere gratitude to B. Leikert, J. Gabel, M. Sing, and R. Claessen from Experimentelle Physik 4 at Physikalisches Institut, at the Würzburg-Dresden Cluster of Excellence ct.qmat, Universität Würzburg, for their contributions to the device fabrication process.

\section[here]{Data Availability Statement}

The theoretical data that supports the findings of this study are available within the article. The experimental data are available from the corresponding author upon reasonable request.


\bibliography{References}

\end{document}